\RequirePackage{fix-cm}
\documentclass[twocolumn,epjc3]{svjour3}  
\smartqed  % flush right qed marks, e.g. at end of proof
\RequirePackage{graphicx}
 \RequirePackage{mathptmx}      % use Times fonts if available on your TeX system
\usepackage{amsmath, amssymb}
\usepackage{color}
\usepackage{ulem}

\renewcommand\sout{\bgroup \color{red} \ULdepth=-.5ex \ULset}
\journalname{Eur. Phys. J. C}
\begin{document}

\title{Net baryon number probability distribution near the chiral phase transition}


\author{Kenji Morita\thanksref{e1,addr1,addr2}
        \and
        Vladimir Skokov\thanksref{addr3,addr7}
 	\and
	Bengt Friman\thanksref{addr4}
	\and
	Krzysztof Redlich\thanksref{addr5,addr6}
	%etc.
}

\thankstext{e1}{e-mail: morita@fias.uni-frankfurt.de}


\institute{Frankfurt Institute for Advanced Studies,
Ruth-Moufang-Str. 1, D-60438 Frankfurt am Main, Germany \label{addr1}
           \and
	   Yukawa Institute for Theoretical Physics, Kyoto University,
	   Kyoto 606-8502, Japan \label{addr2}
           \and
	   Physics Department, Brookhaven National Laboratory, Upton, NY
	   11973, USA \label{addr3}
	   \and
	   Department of Physics, Western Michigan University,
	   1903 W. Michigan Avenue, Kalamazoo, MI 49008, USA \label{addr7}
%           \emph{Present Address:} if needed\label{addr3}
 	    \and
	    GSI, Helmholzzentrum f\"{u}r Schwerionenforschung,
	    Planckstr. 1, D-64291 Darmstadt, Germany \label{addr4}
	    \and
	    Institute of Theoretical Physics, University of Wroclaw,
	    PL-50204 Wroc\l aw, Poland \label{addr5}
	    \and
	    Extreme Matter Institute EMMI, GSI,
	    Planckstr. 1, D-64291 Darmstadt, Germany \label{addr6}
	    }

\date{Received: date / Accepted: date}

\maketitle
\begin{abstract}
 We discuss  properties of the net baryon number probability
 distribution near the chiral phase transition to explore the effect of
 critical fluctuations. Our studies are performed within Landau theory,
 where the coefficients of the polynomial potential are parametrized, so
 as to reproduce the mean-field (MF), the $Z(2)$ and $O(4)$ scaling behaviors of
 the cumulants of the net baryon number.
 We show, that in the critical region, the structure of the probability
 distribution changes, dependently on values of the critical exponents.
 In the MF approach, as well as in the $Z(2)$ universality class,
 the contribution of the singular part of the
 thermodynamic potential tends to broaden the distribution. By contrast,
 in the model with $O(4)$ scaling, the contribution of the singular part
 results in a narrower net baryon number probability distribution  with
 a wide tail.
%\keywords{ \and Second keyword \and More}
\end{abstract}

\section{Introduction}

Fluctuations of conserved charges reflect the critical properties of a
system. In particular, in a strongly interacting medium, fluctuations of
the net baryon number and of the electric charge are valuable probes of
the QCD phase transition
\cite{asakawa,koch,stephanov98:_signat_of_tricr_point_in_qcd,karsch,karschr,hatta03:_proton_number_fluct_as_signal,review}.
Such fluctuations may provide a signature for the conjectured chiral
critical point
\cite{CEP,stephanov09:_non_gauss_fluct_near_qcd_critic_point,stephanov11:_sign_of_kurtos_near_qcd_critic_point},
as well as for the residual criticality of the underlying $O(4)$ transition
\cite{karsch,karschr,braun-munzinger11:_net_proton_probab_distr_in,skokov10:_vacuum_fluct_and_therm_of_chiral_model,sk1,sk2,sk3},
expected in QCD at small densities in the limit of massless $u$ and $d$
quarks \cite{pisarski,karschl}.

Since fluctuations of conserved charges are experimentally accessible
through measurements of their multiplicity distributions,
they have been suggested as probes of the proximity of the chiral crossover
to the freeze-out line in heavy-ion collisions
\cite{karschr,braun-munzinger11:_net_proton_probab_distr_in,Mukherjee}.
A particular role has been attributed to higher order cumulants, which
exhibit an enhanced sensitivity to criticality with increasing order
\cite{karsch,braun-munzinger11:_net_proton_probab_distr_in,sk1,sk2}.
In particular, the sixth- and  higher-order cumulants of the net baryon
number and of the electric charge vary rapidly in the transition
region already at vanishing baryon chemical potential. In the
chiral limit, i.e, for vanishing pion mass, $O(4)$ scaling implies
that e.g. the six order cumulant diverges to $+\infty$ or $-\infty$ when
the critical temperature is approached from below or above, respectively. 
For a small non-zero pion mass, the divergence is replaced by a rapid
temperature dependence, resulting in a change of the sign of the
cumulants close to the chiral crossover transition
\cite{karschr,braun-munzinger11:_net_proton_probab_distr_in,sk2,Mukherjee}.

Consequently, the data on charge fluctuations in high-energy heavy ion
collisions may provide experimental evidence for the chiral
transition of QCD \cite{karschr,Kitazawa}. First measurements of the
multiplicity distribution and corresponding cumulants of the net baryon
number, more precisely the net proton number,
have been obtained  in heavy-ion collisions at
RHIC by the STAR Collaboration \cite{star}. The data
are, at least on a qualitative level, consistent with the residual
criticality, owing to the underlying chiral phase transition. The final
conclusion, however, requires  additional studies of  non-critical 
effects such as e.g. volume fluctuations~\cite{Skokov:2012ds}, acceptance
corrections~\cite{Bzdak:2012ab},  or an  exact baryon number
conservation~\cite{Bzdak:2012an}.

Cumulants of conserved charges have also been studied theoretically. At
small chemical potential $\mu/T \ll 1$, they have been computed 
within lattice QCD~\cite{lgt0,bazavov,fodor,lgtl}, while at large $\mu$,
their properties have been explored within effective chiral
models~\cite{skokov10:_vacuum_fluct_and_therm_of_chiral_model,sk1,sk2,model1,model2,model3,model4,model5},
which share the global symmetries of QCD.

In statistical physics, the $n$-th order cumulant $c_n(T,\mu)$ of a
conserved charge $N$ is obtained by differentiating the thermodynamic
pressure $p(T,\mu)$ with respect to the corresponding chemical potential
$\mu$,
\begin{equation}
 c_n(T,\mu) \equiv \frac{\partial^n [p(T,\mu)/T^4]}{\partial (\mu/T)^n}\,.   \label{cumulants}
\end{equation}
The pressure is related to the grand canonical partition function
$p=(T/V)\ln \mathcal{Z}$.
The cumulants can also be expressed as polynomials in the central
moments $\langle (\delta N)^{n}\rangle$, where
$\delta N = N-\langle N \rangle$.
The $n$-th moment $ \langle N^n \rangle $ is linked to the %normalized
net charge probability distribution
\begin{equation}
 \langle N^n \rangle  = \sum_{N=-\infty}^{\infty}N^n P(N).\label{eq:av_pn}
\end{equation}

Cumulants of the net charge \eqref{cumulants} can exhibit singularities
near the phase transition. Thus, due to criticality, the corresponding
probability distribution \eqref{eq:av_pn} should have the characteristic
shape governed by the universal properties of this transition.

The main objective of this paper is to explore the qualitative features
of the probability distribution of the net baryon number near the chiral
phase transition and their relation to the cumulants. Such studies are
not only of theoretical,  but also of phenomenological interest.	In heavy
ion collisions the probability distribution of conserved charges and
corresponding cumulants are measured to verify the chiral phase
transition or its remnant. In the absence of criticality, different
moments of conserved charges, as well as particle multiplicities,  are
consistent with predictions of the hadron resonance gas model. The
probability distribution of the net baryon number is then governed by the
Skellam
distribution~\cite{braun-munzinger11:_net_proton_probab_distr_in}. Thus,
it is desirable to verify how the critical fluctuations lead to
deviations from	the Skellam distribution.

In this paper we explore the influence of the chiral phase transition on
the probability distribution in the framework of the  Landau theory of
phase transitions. We construct the thermodynamic potential in such a
way, that its regular part results in the Skellam distribution of the net
charge. The singular part is modeled within the Landau theory, where the
coefficients of the polynomial potential are parametrized, so as to
reproduce the mean-field, $O(4)$ and $Z(2)$ scaling behaviors of the
cumulants of the net charge. We quantify the modification of the Skellam
distribution in the presence of the phase transition. We show that the
structure of the probability distribution changes, dependently on the
values of the critical exponents. 
%A characteristic feature of the
%underlying $O(4)$ ($Z(2)$) criticality,  is a narrower (broader)
%probability distribution relative to Skellam function.

In this article, we consider not only expected $O(4)$ universality class, for
simulating chiral QCD, but also $Z(2)$, because this possibility has not been
reliably ruled out yet~\cite{Aoki:2012yj}. Comparison with the $Z(2)$
universality class will also help us to demonstrate significant
properties of the $O(4)$ criticality on the level of the net
charge probability distribution. We consider a system in the chiral
limit in which the critical behavior is most prominent,
 and an analytical treatment of the mean field model is feasible.
The effects of an explicit breaking of the chiral symmetry on the
probability distribution in the $O(4)$ universality
class were recently discussed in Ref.~\cite{pn_qm_frg}.

The paper is organized as follows: in the next Section we introduce the
model for the phase transition. In Section \ref{sec:p_n}, we calculate
the probability distribution and corresponding cumulants near phase
transition. In Section \ref{sec:conclusion}, we present our summary and
conclusions.

\section{Fluctuations in the Landau theory}
\label{sec:landau}

To explore the influence of the second-order phase transition on the
probability distribution of conserved charges, we consider a model based
on the Landau theory of phase transitions. At finite temperature $T$ and
chemical potential $\mu$, the effective	potential density divided by
$T^4$,  $\hat{\omega}(T,\mu;\sigma)=\omega/T^4$, is introduced as a
polynomial in the order parameter $\sigma$, as
\begin{equation}
 \hat{\omega}(T,\mu;\sigma) = \hat{\omega}_{\text{bg}} + \frac{1}{2}a(T,\mu)\sigma^2
  + \frac{1}{4}\sigma^4, \label{eq:landau_omega}
\end{equation}
where $\hat{\omega}_{\text{bg}}$ is a non-singular background
contribution parameterized as

\begin{equation}
 \hat{\omega}_{\text{bg}}=-2 d \cosh(\mu/T).\label{omega_bg}
\end{equation}

The  potential \eqref{eq:landau_omega} exhibits a second order phase
transition located along the critical line $T=T_c(\mu)$, determined by
the condition $a(T,\mu)=0$.   For $d=\pi^4/30$, the characteristic
energy scale of the chiral phase transition in QCD is reproduced. At
$\mu=0$ we set the transition temperature to $T_c=0.15$ GeV.
The precise choice of parameters does not influence our conclusions.

\begin{figure}[!t]
 \includegraphics[width=3.375in]{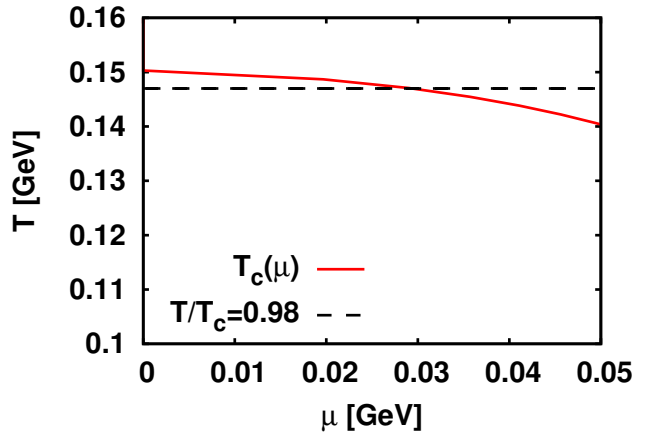}
 \caption{The phase diagram of the Landau model for small chemical
 potentials. The horizontal line indicates the path along which the
 cumulants of baryon number fluctuations are shown in Fig.~\ref{fig:cum_omega}. }
 \label{fig:model_phasediagram}
\end{figure}

\begin{figure*}[t]
 \includegraphics[width=5.75in]{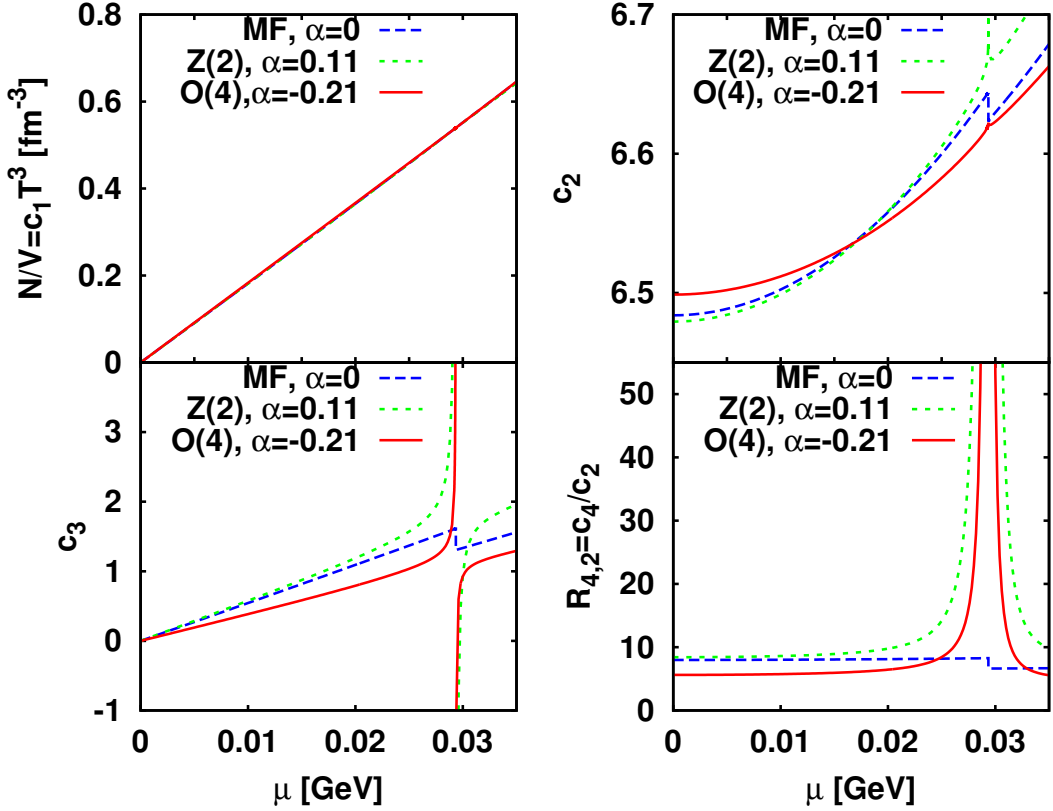}
 \caption{The baryon number density $N/V$ and the first three cumulants
 $c_n$ obtained from the Landau potential  Eqs.~\eqref{eq:omega_reg},
 \eqref{eq:omega_singular} and \eqref{omega_s} for $T/T_{c}=0.98$. The
 dashed lines show  the mean-field (MF) results  with  $\alpha =0$,
 while the solid (dotted) lines correspond to $O(4)$ ($Z(2)$) scaling,
 with $\alpha_{O(4)} =-0.21$ ($\alpha_{Z(2)} =0.11$ ).}
 \label{fig:cum_omega}
\end{figure*}

For a positive $a(T,\mu)$, the minimum of the Landau potential
\eqref{eq:landau_omega} is located at $\sigma=0$, while for $a(T,\mu) <
0$ the order parameter is non-zero, $\sigma=\pm\sqrt{-a(T,\mu)}$ and the
symmetry is spontaneously broken. The parametrization of $a(T,\mu)$ is
chosen, so that the broken symmetry phase is located below the critical
line $T=T_c(\mu)$. Consequently,
\begin{align}
 \hat{\omega}_0 \equiv \hat{\omega}(T > T_c(\mu),\mu)& = \hat{\omega}_{\text{bg}}\label{eq:omega_reg},\\
 \hat{\omega}_1^{\text{MF}} \equiv \hat{\omega}(T < T_c(\mu),\mu)& = \hat{\omega}_{\text{bg}}-\frac{1}{4}|a(T,\mu)|^{2}.
 \label{eq:omega_singular}
\end{align}

For the coefficient $a(T,\mu)$ we employ the following parameterization
\begin{equation}
 a(T,\mu)=  \frac{T-T_c}{T_c}+\cosh\left(\frac{\mu}{T}\right)-1.\label{eq:dtmu}
\end{equation}

For $\mu/T \ll 1$,
Eq.~\eqref{eq:dtmu} reduces to
$a(T,\mu) \simeq A(T-T_c)+B\mu^2\equiv t_{\mu}$,
with $A>0$ and $B>0$. The scaling variable, $t_{\mu}$,  is
frequently used in the literature \cite{karsch,sk3}.

The $\cosh(\mu/T)$ terms in Eqs.~\eqref{omega_bg} and \eqref{eq:dtmu}
account for the periodicity of the thermodynamic potential in an
imaginary chemical potential, $\omega(T,i\theta\, T)$ with $\theta =
\mu_I/T$, which  is a consequence of the $U_B(1)$ symmetry.
We note, that this periodicity is identical to that of QCD
thermodynamic potential \cite{rw}, if $\mu$ corresponds to the baryon
chemical potential. However, since  we do not consider  confinement, the
 $\mu$ is identified as the chemical potential of the elementary fermion.
If one  considers quarks, the thermodynamic potential has a periodicity
$2\pi/N_c$  in an imaginary $\mu$. It also exhibits  a more complicated
temperature dependent structure than that used in Eq. \eqref{eq:dtmu}, when
coupled to gauge fields \cite{immu1,immu2,immu3,morita}.

In the presence of critical fluctuations, the singular part of the
potential scales as $\omega_1 \sim |t_\mu|^{2-\alpha}$ near
$T_c$. Motivated by the
$O(4)$ scaling function \cite{sk2,Engels-Karsch}, we extend the mean
field potential \eqref{eq:omega_singular} by introducing the critical
exponent $\alpha$,

\begin{equation}
 \hat{\omega}_{\text{sing}} =
  -{\text{sgn}(\alpha)\over 4}|a(T,\mu)|^{2-\alpha} \label{omega_s}.
\end{equation}
The total thermodynamic potential is then given by 
\begin{equation}
\hat{\omega}_1 = \hat{\omega}_{0} + \hat{\omega}_{\text{sing}}.\label{eq:omega_total}
\end{equation}

The exponent $\alpha$  in Eq.~\eqref{omega_s} is introduced to
parametrize the scaling properties of the thermodynamic potential
in the critical region. 
The sgn($\alpha$) is introduced to reproduce the scaling function also for 
$\alpha < 0$.
Thus, by tuning $\alpha$, we can switch between $O(4)$, $Z(2)$
and mean-field
scaling~\footnote{This admittedly {\em ad hoc} procedure provides
a transparent framework for exploring the effect of critical scaling on
the cumulants of the net baryon number.}.
For $\alpha=0$, the singular part of $\omega$ exhibits mean-field
scaling. On the other hand, for $\alpha\simeq -0.21$
($\alpha\simeq 0.11$),
the Landau potential emulates the critical behavior
of the $O(4)$ ($Z(2)$)
spin system in 3-dimensions~\footnote{Since other critical exponents are not reproduced, this statement
is limited to quantities governed by the critical exponent $\alpha$.}.

Figure \ref{fig:model_phasediagram} shows the critical line $T_c(\mu)$
of the second order transition obtained in the Landau theory, using the
parametrization \eqref{eq:dtmu}.
The behavior of the first four cumulants along the line of constant
temperature $T/T_c=0.98$ for different $\mu$ is illustrated in
Fig.~\ref{fig:cum_omega}. Their critical properties can be quantified
from the singular potential \eqref{omega_s}.

Close to the critical point,
$|T-T_{c}|/T_{c}\ll 1$,
the singular part of the potential scales as $\omega_1 \sim |t_\mu|^{2-\alpha}$.
Consequently, at  $\mu=0$, only even order cumulants of the net baryon
number are finite and their singular part scales as
\[
 c_{2k}^{\text{sing}} \sim |t_\mu|^{2-\alpha-k}.
\]
Thus, for  the $O(4)$ exponent $\alpha\simeq -0.21$, the first
divergent cumulant is of  the sixth order \cite{karschr,sk2},
while for the $Z(2)$ it is of the fourth order.

For $\mu\neq 0$,  the leading singularities of  cumulants
\begin{equation}
c_n^{\text{sing}} \sim \mu^n|t_\mu|^{2-\alpha-n},
\end{equation}
imply, that for $O(4)$ ($Z(2)$)  %  $\alpha\simeq -0.21$,
only  cumulants  with $n\geq 3$ ($n\geq 2$) diverge at the critical
point. 

With the mean-field critical exponent $\alpha=0$, the singular part of
the  cumulants remains finite near the critical point. Indeed,
differentiating Eq.~\eqref{eq:omega_singular}, one finds, that for $T<T_c$,
%the non-vanishing  cumulants are given by
\begin{alignat}{3}
c_1^{\text{sing}}&=-B\mu |t_\mu|,\quad &c_2^{\text{sing}}&=  -B|t_\mu|+2B^2\mu^2, \\
c_3^{\text{sing}}&=6B^{2}\mu, &c_4^{\text{sing}}&= 6B^2.
\end{alignat}
Thus, in the mean-field approach, the cumulants are finite and in
general, discontinuous at the critical point.

The singular behavior of the cumulants shown in Fig.~\ref{fig:cum_omega}
and their dependence on the value of the critical exponent, must be also
reflected in the corresponding probability distributions $P(N)$ of the
net baryon number. In the following, we compute $P(N)$ within the mean-field, 
the $O(4)$ and the $Z(2)$  parametrization of the thermodynamic
potential,  to explore the characteristic features of $P(N)$  which are
responsible for the critical  behavior of the cumulants seen in
Fig.~\ref{fig:cum_omega}.

\section{The probability distribution of the net  baryon  number}
\label{sec:p_n}

Consider a grand canonical thermodynamic system  of
charged particles $q$ and anti-particles $\bar{q}$ at volume $V$,
temperature $T$ and at   chemical potential $\mu$. The latter is
related to the conserved net charge $N=N_q-N_{\bar q}$.

The normalized probability distribution $P(N)$ to find
$N$ net charges in  volume $V$,  is given in terms of
the canonical $Z(T,V,N)$ and the grand canonical $\mathcal{Z}(T,V,\mu)$
partition function \cite{hwa,cleymans1,turko},
\begin{equation}
 P(N;T,\mu,V) = \frac{Z(T,V,N)e^{\beta \mu N}}{\mathcal{Z}(T,V,\mu)},\label{eq:pN}
\end{equation}
where $\beta=1/T$.

The partition functions are related by the fugacity expansion,
\begin{equation}
 \mathcal{Z}(T,V,\mu) = \sum_{N}\lambda^N Z(T,V,N)\label{eq:fug_exp},
\end{equation}
where $\lambda=e^{\beta \mu}$ is the fugacity parameter.

\begin{figure}[t]
 \includegraphics[width=0.40\textwidth]{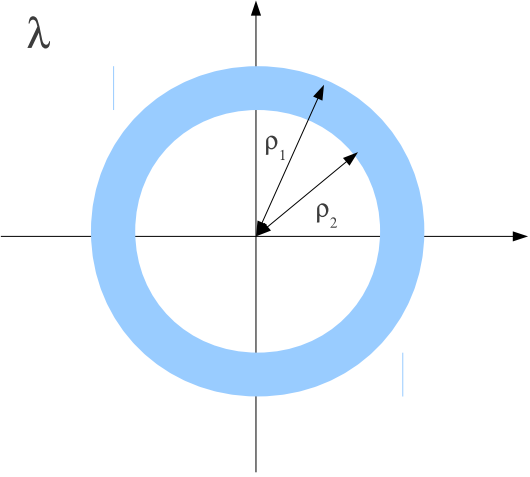}
 \caption{The integration contour $C$ in Eq.~\eqref{eq:cano} lies within
 a singularity free annulus in the complex $\lambda$-plane, enclosing
 the origin.}
 \label{fig:laurent1}
\end{figure}

Thus, $Z(T,V,N)$ is just the $N$-th order coefficient in the Laurent
expansion of $\mathcal{Z}(T,V,\mu)$ around $\lambda=0$. Consequently,
 $Z(T,V,N)$ can be obtained from
\begin{equation}
 Z(T,V,N) = \frac{1}{2\pi i }\oint_C d\lambda
	\frac{\mathcal{Z}(T,V,\mu)}{\lambda^{N+1}},\label{eq:cano}
\end{equation}
where the integration contour $C$ must lie in an annulus enclosing the
origin in the complex $\lambda$-plane. Inside the annulus the integrand
$\mathcal{Z}(T,V,\mu)$ must be analytic (see Fig.~\ref{fig:laurent1}).

Taking $C$ on the unit circle, $\lambda=e^{i\theta}$, one finds the
well-known result \cite{hwa,hagedorn},
\begin{equation}
 Z(T,V,N) = \frac{1}{2\pi}\int_{0}^{2\pi}d\theta e^{-i\theta N} \mathcal{Z}(T,V,i\, T\, \theta).\label{eq:cano_theta}
\end{equation}
We note, that Eq.~\eqref{eq:cano_theta} holds  only when
$\mathcal{Z}(T,V,\mu)$ is analytic on the unit circle. In general, the
partition function for finite systems~\footnote{In a finite system, the
partition function has Yang-Lee zeroes at complex $\mu$, which in the
thermodynamic limit turns into
cuts~\cite{stephanov06:_qcd_critic_point_and_compl}, with branch points
at the critical point and at $|\mu|=\infty$.}, e.g. in lattice QCD
simulations in a finite volume \cite{alford99:_imagin_chemic_poten_and_finit,forcrand06:_finit_densit_qcd_with_canon_approac,ejiri08:_canon_qcd,li:_finit_qcd_n_n,li11:_critic_point_of_n_f,XQCDJ},
is an analytic function of $\mu$. Hence, in this case,  the integral
\eqref{eq:cano} %and \eqref{eq:cano_theta}
is well defined and independent of the radius of the contour.

However, for the partition function $\mathcal{Z}=e^{-\beta V \omega}$,
where $\omega=\hat{\omega}T^4$ in Eqs.~\eqref{eq:omega_reg},
\eqref{eq:omega_singular} and \eqref{eq:omega_total} is approximated by the thermodynamic
potential density in the {\it thermodynamic limit}, while $V$ is kept to be
finite, there are singularities in the complex $\mu$ plane,
which must be properly accounted for.
This approximate treatment is justified for $|N|\ll N^*$, where $N^*$ is a
characteristic of the partition function of a finite system in the
Yang-Lee theory of phase transitions~\footnote{In the original work of
C. N. Yang and T.D. Lee~\cite{Yang:1952be}, $N^*$ is the maximum number
of classical particles that can be packed into the volume $V$. The
Yang-Lee theory was later extended to a more general class of models.}.

In general, to determine $N^*$ for a given system, a microscopic
computation of the grand canonical partition function in a finite volume
is needed. However, such a calculation cannot be carried out within the
Landau model considered here. Consequently, in the following, we assume
that $N^*$ is larger than the maximal $N$, needed in the numerical
evaluation of the probability distribution $P(N)$.
We shall discuss the properties of the canonical partition function under
this assumption.

In the chiral limit and for $T < T_c$, a critical point exists on the positive real
fugacity axis, $\lambda=\lambda_c > 1$,~\cite{skokov11:_mappin}. Charge
conjugation symmetry of the partition function, $\mu \rightarrow -\mu$,
implies,  that there is a corresponding critical point at
$\lambda=1/\lambda_c$. This symmetry is respected also by the
thermodynamic potential through \eqref{omega_bg} and \eqref{eq:dtmu}.

At fixed temperature, the critical points at $\mu=\pm\mu_c$, or
equivalently at $\lambda= \lambda_c^{\pm 1}$,  are
branch point singularities of the order parameter $\sigma$ in the
complex $\mu$ (or $\lambda$) plane~\cite{stephanov06:_qcd_critic_point_and_compl,Friman:2012gg}, with the
critical exponent $\beta$ since $\sigma \sim (-t_{\mu})^{\beta}$.

In the mean-field case $\beta=\frac12$, thus the singularity is of the
square root type. In addition,  since 
$\hat\omega_{\text{sing}} \sim |a(T,\mu)|^2 \,\theta(-a(T,\mu))$
 and $a\sim t_{\mu}$, the
singularity of the  thermodynamic potential at the critical point
is a discontinuity in the higher derivatives,  i.e. in the cumulants
$c_{n}$ for $n\geq 2$, as seen in Fig.~\ref{fig:cum_omega}.

In the $O(4)$ {and $Z(2)$} case, the critical points are branch
points also of the thermodynamic potential,
$\hat\omega_{\text{sing}} \sim |t_{\mu}|^{2-\alpha}$.
Consequently, the thermodynamic potential has cuts originating at the
branch points. A convenient choice is to place the cuts on the real
axis, between the critical points $\mu=\mu_c$ and $\mu=\infty$, as well
as between $\mu=-\mu_c$ and $\mu=-\infty$.  The corresponding cuts in
the fugacity are located between $\lambda=\lambda_c >1$ and
$\lambda = \infty$ and between $\lambda=1/\lambda_c$ and $\lambda=0$,
respectively \cite{skokov11:_mappin}. As a result, in the thermodynamic
limit, the thermodynamic potentials of different phases (in the present
case, the $\omega_0$ and $\omega_1$) correspond to different Riemann
surfaces connected through cuts~\cite{Friman:2012gg}.

If the grand canonical partition function $\mathcal{Z}$ exhibits branch
singularities, then the coefficients of the Laurent expansion
$Z(T,V,N)$,  depend on the integration contour in
Eq.~\eqref{eq:cano}. In the broken-symmetry phase, the annulus is
singularity free for $\rho_1 = \lambda_c >1$ and
$\rho_2 =1/\lambda_c <1$ (cf.\ Fig.~\ref{fig:laurent1}). Thus, the
coefficients of the Laurent expansion corresponding to the
broken-symmetry phase, are obtained from \eqref{eq:cano_theta}
with $ \mathcal{Z}= e^{-\beta V \omega_1}$.
On the other hand, the partition function
$\mathcal{Z} = e^{-\beta V\omega_0}$
of the symmetric phase is singularity free outside the
annulus of the broken symmetry phase. Hence, the corresponding Laurent
coefficients are obtained by integrating \eqref{eq:cano} along e.g. a
circular contour with radius $\rho < 1/\lambda_c$ or $\rho > \lambda_c$.

Thus, in general, for a given $N$  we obtain two competing partition
functions. Obviously, the partition function of a thermodynamic system
must be unique. The reason for this ambiguity is that we approximate
$\omega$ in $\mathcal{Z}$ by the grand canonical thermodynamic potential
density in the thermodynamic limit,  while the system is kept in a finite
volume $V$. { If $V$ in the thermodynamic potential
density is kept finite when the canonical partition function is
calculated,}
the uniqueness of the canonical partition function is restored,
{ even after taking the thermodynamic limit}. This is because,  the  solution
corresponding to a larger value of $\omega$ is suppressed.

In the following, we consider only the canonical partition function
obtained from the grand canonical one,  using Eq. \eqref{eq:cano} for
$\mu < \mu_c$, i.e. for the broken symmetry phase. The relation between
the probability distribution $P(N)$ and the canonical partition function
\eqref{eq:pN} implies, that the structure of $P(N)$ is entirely governed
by the properties of $Z(T,V,N)$.

\subsection{Probability distribution of the nonsingular potential}

The probability distribution corresponding to the nonsingular part of
the thermodynamic potential density $\omega_0$  \eqref{omega_bg} can be computed analytically.
Indeed, using the generating function of the modified Bessel function $I_n(x)$,
\begin{equation}
 e^{\frac{x}{2}(\lambda+\frac{1}{\lambda})} = \sum_{n=-\infty}^{\infty}I_n(x)\lambda^n,\label{bessel}
\end{equation}
one can directly expand the grand canonical partition function
$\mathcal{Z}=e^{-\beta V \omega_{0}}$, without passing through the
integral representation  \eqref{eq:cano}, and obtain
\cite{hwa,cleymans2}
\begin{equation}
 Z_{0}(T,V,N) = I_N(2d VT^3).
\end{equation}

The probability distribution from  the nonsingular part of the Landau
potential \eqref{eq:omega_reg}  is then obtained from Eq.~\eqref{eq:pN}
as the Skellam function,
\begin{equation}
 P^{\rm NS}(N;T,V,\mu) = I_N(2dVT^3)e^{(\mu N + \Omega_0)/T},\label{eq:pn_nonsingular}
\end{equation}
where $\Omega_0 = V \omega_{0}=-2dVT^4 \cosh(\mu/T)$ is the
corresponding thermodynamic potential.

\begin{figure}[!t]
 \includegraphics[width=3.375in]{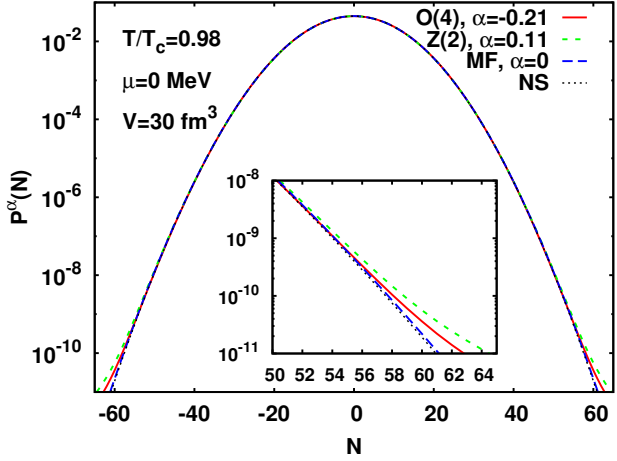}
 \caption{Probability distributions of the net baryon number in
 the Landau model. The dashed line shows the singular case with the mean
 field exponent $\alpha=0$, from Eq.~\eqref{pmf}. Solid and dotted lines denote the $O(4)$
 and the $Z(2)$ critical case in which $P^\alpha(N)$ is
 obtained numerically for $\alpha=-0.21$ and $0.11$, respectively.
 The nonsingular distribution $P^{\rm NS}(N)$ from Eq.~\eqref{eq:pn_nonsingular}
 is plotted as the thin dotted line.
 The calculations were done at $\mu=0$ and for  $V=30$ fm$^3$.}
 \label{fig:pn}
\end{figure}

\subsection{Probability distribution of the mean-field potential}

For the Landau potential \eqref{eq:omega_singular} with the mean-field
exponent $\alpha=0$, the canonical partition function $Z(T,V,N)$ can
also be obtained analytically, using the same procedure as was discussed
above.

The partition function for the singular case reads
\begin{align}
 e^{-\beta V\omega_1}&= \exp\left\{ VT^3 \left[d \left(\lambda +
						 \frac{1}{\lambda}\right)+\frac{1}{4}a^2(T,\mu)\right]\right\}.\label{eq16}
\end{align}
By expanding the argument of the exponential
\begin{align}
 |a(T,\mu)|^2& = (t+1)^2+\frac{1}{2}-(t+1)\left( \lambda+\frac{1}{\lambda} \right)
 + \frac{1}{4}\left( \lambda^2 + \frac{1}{\lambda^2} \right),
\end{align}
 with $t=1-T/T_c$ and using Eqs.~\eqref{bessel} and \eqref{eq:cano_theta} one finds
\begin{align}
 Z_{c}^{\text{MF}}(T,V,N)&= e^{\frac{VT^3}{4}[(t+1)^2+\frac{1}{2}]} \nonumber\\
	\times \sum_{\ell=-\infty}^{\infty}I_{N-2\ell} & [(2d-(t+1)/2)VT^3]I_\ell\left(\frac{VT^3}{8}\right)\label{eq:zc_MF}.
\end{align}
The probability distribution $P^{\rm MF}(N)$ can be then obtained from
Eqs.~\eqref{eq16},  \eqref{eq:zc_MF} and \eqref{eq:pN},  as
\begin{align}
 P^{\text{MF}}(N;T,V,\mu)&= e^{VT^3\left( \frac{t+1}{4}-d \right)
	\left( \lambda + \frac{1}{\lambda} \right)-\frac{1}{16}\left(
	 \lambda^2+ \frac{1}{\lambda^2}\right) +\frac{\mu N}{T}} \nonumber\\
	\times \sum_{\ell=-\infty}^{\infty}I_{N-2\ell} & [(2d-(t+1)/2)VT^3]I_\ell\left(\frac{VT^3}{8}\right)\label{pmf}.
\end{align}

Figure \ref{fig:pn} shows the resulting probability distributions
calculated at $\mu=0$ for $V=30$ fm$^3$. The mean-field probability
distribution $P^{\rm MF}(N)$ is broader than  $P^{\rm NS}(N)$
for $|N|>30$. This feature is particularly  evident in the ratio
$P^{\rm MF}(N)/P^{\rm NS}(N)$,   shown  in Fig.~\ref{fig:pn_ratio}.
%\corr{Furthermore, the detailed structure around $N=0$ indicates the
%peak of the distribution is shapened. }
Thus, for the Landau potential with $\alpha=0$,  the
criticality is expressed in the probability distribution as broadening
of the regular Skellam function.\footnote{In Ref.~\cite{pn_qm_frg}, the mean
field calculation shows a narrower distribution than the Skellam
distribution. However, as discussed in Ref.~\cite{pn_qm_frg}, this is
caused by Fermi statistics, which is not accounted for in the present
model.}

\subsection{Probability distribution of the $O(4)$ and $Z(2)$ potentials}

With the mean-field exponent $\alpha=0$, the net charge probability
distributions were  calculated exactly. However, such an exact
derivation cannot be applied for the singular part with
$\alpha\neq 0$.
In this case, one can only use the numerical or some approximate
analytical methods to obtain the canonical partition function and the
corresponding probability distribution $P^\alpha(N)$.
\begin{figure}[!t]
 \includegraphics[width=3.375in]{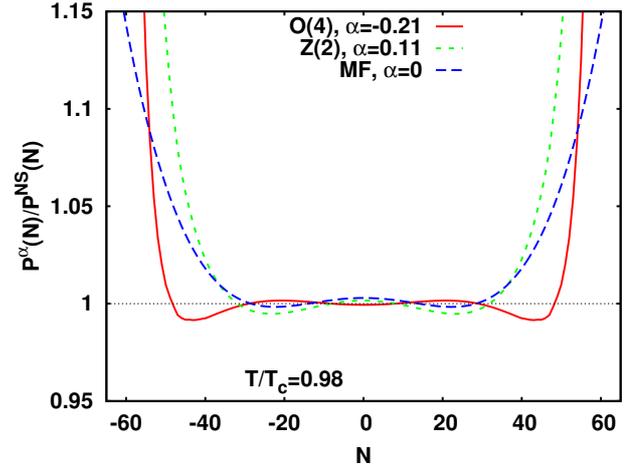}
 \caption{The ratios of the distributions from Fig.~\ref{fig:pn} to the
 non-singular probability distribution $P^{\rm NS}(N)$ from
 Eq.~\eqref{eq:pn_nonsingular}.} \label{fig:pn_ratio}
\end{figure}

The most direct approach to get $Z(T,V,N)$, is the numerical
integration of Eq.~\eqref{eq:cano_theta}.
There are, however, limitations of this method, in particular at large
$N$, due to the oscillatory structure of the integrand.
We adopt the \texttt{qawo} subroutine in QUADPACK package library for
computing oscillatory integrals and check convergence within $1\%$ accuracy.

To test the numerical method, we first compare the probability
distribution  $P^{\rm MF}(N)$ obtained analytically in
Eq.~\eqref{pmf} with the numerical integration of
Eq.~\eqref{eq:cano_theta}.
		%Figure~\ref{fig:benchmark} shows the ratio
		%of the resulting probability distributions.
The numerical integration reproduces the analytical result up to
$|N| \simeq 60$,  at which $P(N)\sim 10^{-12}$,
but fails for larger $N$ { due to the round-off error}. Thus, to
calculate  $P^\alpha(N)$ for the  model with
nontrivial $\alpha$,  one can trust the numerical methods
% integration can be
%used with  a confidence also
only up to some finite value of $|N|$,  where  $P(N)\sim 10^{-12}$.

The numerical results for $P^\alpha(N)$ are shown in Fig.~\ref{fig:pn}.
In Fig.~\ref{fig:pn_ratio}, the same probability distributions are
normalized to the non-singular distribution $P^{\rm NS}(N)$, 
given in Eq.~\eqref{eq:pn_nonsingular}. For $\alpha=-0.21$ the ratio
$P^{\alpha}(N)/P^{\rm NS}(N)$  differs considerably from that of the
mean-field model. The ratio $P^{\rm MF}(N)/P^{\rm NS}(N)$
increases with 
%\footnote{\sout{In this discussion
%we ignore the very small differences that appear for $|N|\lesssim 30$.}}
$|N|$ for $|N| > 30$ and exhibits oscillations about unity for $|N|$.
While the ratio $P^{O(4)}(N)/P^{\text{NS}}(N)$ also oscillates for small
$|N|$, albeit with the  opposite phase,
it then decreases up to an intermediate value of $|N| \simeq 40$, and
then increases sharply. This indicates that for the critical exponent
$\alpha\simeq -0.21$, the probability distribution, relative to the
Skellam distribution, is  narrower for the intermediate, and  wider for larger
values of $|N|$. With the $Z(2)$ exponent $\alpha\simeq 0.11$, the ratio is very similar
to the MF one for $|N| < 40$, while at larger $|N|$ it shows a
much sharper increase than either the MF and $O(4)$ distribution. These
properties are reflected also in the critical behavior of the cumulants.

The canonical partition function $Z(T,V,N)$ can also be computed by applying the
method of the steepest descent to the integral in Eq.~\eqref{eq:cano}.
The region of applicability of this method is found, however,  to be
limited to quite small $N\sim 10$,  for the same set of parameters as
in Fig.~\ref{fig:pn}, because of the analytic structure of the integrand.

\begin{figure}[ht]
 \includegraphics[width=3.05in]{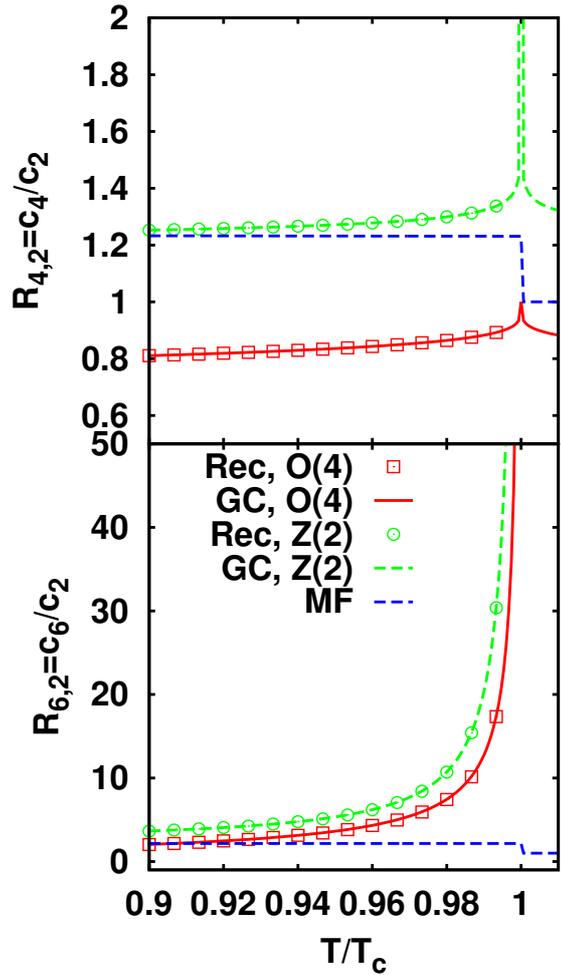}
 \caption{Cumulants at finite temperature and at $\mu=0$ calculated
 from the probability  distribution $P^{\alpha}(N)$ with
 $\alpha_{O(4)}=-0.21$ (boxes) and $\alpha_{Z(2)}=0.11$ (circles). 
 The $P^\alpha(N)$ was calculated numerically up to  $|N|=67$. The curve
 labeled GC is obtained directly from the grand canonical partition
 function using Eq.~\eqref{cumulants}.}
\label{fig:c6mu0}
\end{figure}

\subsection{Reconstructing cumulants from the probability distributions}
We have argued, that the structure of the net charge probability
distribution is sensitive to the phase transition and its properties.
To verify if the observed
characteristic structures of $P(N)$ are entirely due to criticality,
one would need to check, whether the cumulants reconstructed from these
distributions have properties expected near the critical point.
We calculate different cumulants

\begin{align}\label{eq:c2}
 c_2(T,\mu) &= \frac{1}{VT^3}\langle (\delta N)^2 \rangle,\\
 c_3(T,\mu) &= \frac{1}{VT^3}\langle (\delta N)^3 \rangle,\\
 c_4(T,\mu) &= \frac{1}{VT^3}\left[ \langle (\delta N)^4 \rangle -3
 \langle (\delta N)^2 \rangle^2 \right], \label{eq:c4}\\
 c_6(T,\mu)&= \frac{1}{VT^3}\left[ \langle (\delta N)^6 \rangle -15 \langle (\delta N)^4 \rangle
 \langle (\delta N)^2 \rangle\right. \nonumber\\
	&\left.- 10 \langle (\delta N)^3 \rangle^2 +30\langle (\delta N)^2
 \rangle^3 \right]\label{eq:c6},
\end{align}
directly from the probability distributions $P(N)$ following Eq.~\eqref{eq:av_pn}.

The analytic form of the mean-field  distribution $P^{\rm MF}(N)$,
fully  reproduces  all properties of cumulants at vanishing and finite $\mu$.
The approximate $P^{\rm MF}(N)$ obtained by the numerical integration,
yields cumulants $c_n$, which are consistent with exact results, except
for some deviations in the immediate neighborhood of the phase
transition,  at a non-vanishing chemical potential.

In the model with the $O(4)$  exponent,
%\corr{As shown in Fig.~\ref{fig:c6mu0},
the  $P^\alpha(N)$  calculated  numerically up to $|N| \leq 67$,  at
$\mu=0$  and $T/T_c=0.99$, reproduces the exact results of cumulants in
the vicinity of the critical point, as seen in
Fig.~\ref{fig:c6mu0}. This is not only the case for the non-critical
second- and fourth-order cumulants,  but also for the critical sixth-order
cumulant, which diverges at the critical point. 
Similarly, one observes that the divergent behavior of $c_4$ and $c_6$ for
the $Z(2)$ exponent is  also well reproduced by the corresponding probability
distribution.\footnote{
We note, that at finite chemical potential a multiplicative factor
$e^{\beta \mu N}$ in Eq.~\eqref{eq:pN} enhances the  significance of
large $N$-contributions to $P(N)$.    
This causes  difficulties to  reproduce the  critical structure of a
higher order cumulants at finite $\mu$ directly from the numerically
calculated  probability distribution.
}

Since both the $O(4)$ and $Z(2)$ models exhibit the positively divergent
$c_6$ at $\mu=0$, the long tail in the probability distribution can be
attributed to this divergence. On the other hand, the difference of the
$O(4)$ and $Z(2)$ probability distributions up to $|N|\sim 40$ shown in
Fig.~\ref{fig:pn_ratio} can be understood as a consequence of the
different sign of the critical exponent. In the chiral limit, the
contribution of the singular part of
the thermodynamic potential to the fourth order cumulant is negative 
and vanishes at the critical temperature \cite{sk2}.
This means that, for $\alpha < 0$, $c_4$ is smaller than the Skellam
counterpart for $T\neq T_{c}$. This accounts for the narrowing around
$|N|\sim 40$ shown in Fig.~\ref{fig:pn_ratio}.

These results confirm our main conclusion, that the characteristic
behavior of the net charge probability distributions in the $O(4)$ and
$Z(2)$ symmetric models near the critical temperature, is a reflection of the 
corresponding second order phase transition.

\section{Conclusions}
\label{sec:conclusion}
We have explored the influence of the second order chiral phase
transition on the properties of the probability distribution $P(N)$ of
the net baryon number.

We have modeled the critical behavior within the Landau theory of phase
transitions, where the singular part of an effective thermodynamic
potential is expressed as a polynomial in the order parameter.
The coefficients of the potential were  parameterized,  so as to reproduce the
 mean-field, the $O(4)$ and  the $Z(2)$ scaling behavior of cumulants of the net
charge. The structure of the regular part of the potential was adopted
from the hadron resonance gas model to reproduce the Skellam function
for the net-baryon number probability distribution.

In the mean-field approach, the probability distribution was computed
analytically. In the model which reproduces the $O(4)$ and $Z(2)$ scaling of the
net charge fluctuations, the probability distribution was obtained
numerically.

We have shown, that the shape of the distribution depends on the
universality class of the second-order phase transition. 
In particular, with the mean field and $Z(2)$ critical
exponents, the probability distribution $P(N)$ is characterized by
a somewhat sharper peak and broader tail, compared to the non-singular
Skellam function. On the other hand, in the $O(4)$ model,  
the distribution is less peaked but narrower for intermediate values of
$|N|$, followed by a wider tail for large $|N|$.

We have explicitly demonstrated, at a vanishing chemical potential, that
the above behaviors of the  net baryon number probability distributions
indeed capture the critical properties of a system with the $O(4)$ and
$Z(2)$ scaling. This has been done by reconstructing the critical
structure of the higher order cumulants directly from these
distributions.

Although our model  approach does not provide a quantitative description
of the probability distribution relevant for QCD and heavy ion
phenomenology, the qualitative aspects of our results on
the influence of criticality on the probability distribution and its
properties are nevertheless of general validity in the chiral
limit. Effects of
an explicit breaking of the chiral symmetry, together with a more consistent
treatment of the critical fluctuations, have been recently examined in
\cite{pn_qm_frg}.
These results indicate, that 
a direct comparison of measured probability distribution of conserved
charges obtained in heavy-ion collisions with a non-critical
distribution of an ideal gas, e.g. the Skellam function, can provide
pertinent information on the underlying chiral phase transition of QCD 
and its origin.

\begin{acknowledgements}
The authors gratefully acknowledge F.~Karsch for discussions,
 particularly on the $O(4)$ scaling function.
K.M. would like to thank RIKEN-BNL Research Center for their warm
hospitality during his visit where part of this work was completed.
He also would like to thank P.~de Forcrand and A.~Ohnishi for fruitful
 discussions.
This work was in part supported by HIC for FAIR, by the young
 researchers exchange program by
the Yukawa Institute for Theoretical Physics, by  Yukawa International
Program for Quark-Hadron Sciences at Kyoto University and by the
Grants-in-Aid for Scientific Research from JSPS No.24540271.
K.R. acknowledges partial support of the Polish Ministry of National
Education (MEN).
The research of V.S. is supported under Contract No. DE-AC02-98CH10886
with the U. S. Department of Energy.
B.F. is supported in part by the ExtreMe Matter Institute EMMI.
\end{acknowledgements}


\begin{thebibliography}{15}
 \bibitem{asakawa}
	 M.~Asakawa, U.~W.~Heinz and B.~Muller,
	 %``Fluctuation probes of quark deconfinement,''
	 Phys.\ Rev.\ Lett.\  {\bf 85}, 2072 (2000); Nucl.\ Phys.\ {\bf
	 A698}, 519 (2002).
	
 \bibitem{koch}
	 S.~Jeon and V.~Koch,
	 %``Charged particle ratio fluctuation as a signal for QGP,''
	 Phys.\ Rev.\ Lett.\  {\bf 85}, 2076 (2000);
	 V.~Koch, M.~Bleicher and S.~Jeon,
	 %``Event-by-event fluctuations and the QGP,''
	 Nucl.\ Phys.\ {\bf A698}, 261 (2002);
	 V.~Koch,
	 %``Correlations and fluctuations: status and perspectives,''
	 J.\ Phys.\ G {\bf 35}, 104030 (2008).
	
 \bibitem{stephanov98:_signat_of_tricr_point_in_qcd}
	 M.~Stephanov, K.~Rajagopal and  E.~Shuryak,
	 Phys.\ Rev.\ Lett. \textbf{81}, 4816 (1998);
	 Phys.\ Rev.\ D \textbf{60}, 114028 (1999).
						
\bibitem{karsch}
	S.~Ejiri, F.~Karsch and K.~Redlich,
	%``Hadronic fluctuations at the QCD phase transition,''
	Phys.\ Lett.\ B {\bf 633}, 275 (2006).

\bibitem{karschr}
	F.~Karsch and K.~Redlich,
	%``Probing freeze-out conditions in heavy ion collisions with moments of charge fluctuations,''
	Phys.\ Lett.\ B {\bf 695}, 136 (2011)

 \bibitem{hatta03:_proton_number_fluct_as_signal}
	 Y.~Hatta and M.~A. Stephanov,
	 Phys.\ Rev.\ Lett. \textbf{91}, 102003 (2003).

 \bibitem{review}
	 K. Fukushima, and C. Sasaki,  Prog.\ Part.\ Nucl.\
	 Phys. \textbf{72}, 99 (2013).

 \bibitem{CEP}
	 M. Asakawa, and  K. Yazaki,  Nucl.\ Phys.\ {\bf A504}, 668 (1989).

 \bibitem{stephanov09:_non_gauss_fluct_near_qcd_critic_point}
	 M.~A. Stephanov,
	 Phys.\ Rev. Lett. \textbf{102}, 032301 (2009).
	 
 \bibitem{stephanov11:_sign_of_kurtos_near_qcd_critic_point}
	 M.~A. Stephanov,
	 Phys.\ Rev.\ Lett. \textbf{107}, 052301 (2011).

 \bibitem{braun-munzinger11:_net_proton_probab_distr_in}
	 P.~Braun-Munzinger, B.~Friman, F.~Karsch, K.~Redlich and V.~Skokov,
	 Phys.\ Rev.\ C \textbf{84},064911 (2011); Nucl.\ Phys.\ {\bf
	 A880}, 48 (2012).

 \bibitem{skokov10:_vacuum_fluct_and_therm_of_chiral_model}
	 V.~Skokov, B.~Friman, E.~Nakano, K.~Redlich and B.-J. Schaefer,
	 Phys.\ Rev.\ D \textbf{82}, 034029 (2010).

 \bibitem{sk1}
	 V.~Skokov, B.~Friman, F.~Karsch and K.~Redlich,
	 %``Charge fluctuations in chiral models and the QCD phase transition,''
	 J.\ Phys.\ G  {\bf 38}, 124102 (2011).

 \bibitem{sk2}
	 B.~Friman, F.~Karsch, K.~Redlich and V.~Skokov,
	 %``Fluctuations as probe of the QCD phase transition and freeze-out in heavy ion collisions at LHC and RHIC,''
	 Eur.\ Phys.\ J.\ C {\bf 71}, 1694 (2011).

 \bibitem{sk3}
	 V.~Skokov, B.~Friman, and K.~Redlich, 
	 Phys.\ Rev.\ C {\bf 83}, 054904 (2011).

 \bibitem{pisarski}
	 R.~D.~Pisarski and F.~Wilczek,
	 %``Remarks on the Chiral Phase Transition in Chromodynamics,''
	 Phys.\ Rev.\ D {\bf 29}, 338 (1984).

 \bibitem{karschl}	
	 S.~Ejiri, F.~Karsch, E.~Laermann, C.~Miao, S.~Mukherjee,
	 P.~Petreczky, C.~Schmidt, W.~Soeldner and W.~Unger, 
	 Phys.\ Rev.\ D {\bf 80},  094505 (2009).
	 
 \bibitem{Mukherjee}
	 F.~Karsch, E.~Laermann, C.~Miao, S.~Mukherjee, P.~Petreczky,
	 C.~Schmidt, W.~Soeldner and W.~Unger,
	 %``The phase boundary for the chiral transition in (2+1)-flavor QCD at small values of the chemical potential,''
	 Phys.\ Rev.\ D {\bf 83}, 014504 (2011).
	 
 \bibitem{Kitazawa}	
	 M.~Kitazawa and M.~Asakawa,
	 %``Revealing baryon number fluctuations from proton number fluctuations in relativistic heavy ion collisions,''
	 Phys.\ Rev.\ C {\bf 85}, 021901 (2012).
	 
 \bibitem{star}
	 M.~M.~Aggarwal {\it et al.} [STAR Collaboration],
	 Phys.\ Rev.\ Lett.\  {\bf 105}, 022302 (2010);
	 X.~Luo, {\it et al.}, [for the STAR Collaboration],
	 J.~Phys.~Conf.~Ser. \textbf{316} 012003 (2011).

 \bibitem{Skokov:2012ds}
	 V.~Skokov, B.~Friman and K.~Redlich,
	 %``Volume Fluctuations and Higher Order Cumulants of the Net Baryon Number,''
	 arXiv:1205.4756 [hep-ph].
	 %%CITATION = ARXIV:1205.4756;%%

 \bibitem{Bzdak:2012ab}
	 A.~Bzdak and V.~Koch,
	 %``Acceptance corrections to net baryon and net charge cumulants,''
	 Phys.\ Rev.\ C {\bf 86}, 044904 (2012)
	 %[arXiv:1206.4286 [nucl-th]].
	 %%CITATION = ARXIV:1206.4286;%%
	 %4 citations counted in INSPIRE as of 15 May 2013
	
	 %\cite{Bzdak:2012an}
 \bibitem{Bzdak:2012an}
	 A.~Bzdak, V.~Koch and V.~Skokov,
	  %``Baryon number conservation and the cumulants of the net proton distribution,''
	  Phys.\ Rev.\ C {\bf 87}, 014901 (2013)
	 %[arXiv:1203.4529 [hep-ph]].
	  %%CITATION = ARXIV:1203.4529;%%
	 %7 citations counted in INSPIRE as of 15 May 2013

 \bibitem{lgt0}
	 C.~R.~Allton, S.~Ejiri, S.~J.~Hands, O.~Kaczmarek, F.~Karsch,
	 E.~Laermann and C.~Schmidt,
	 Phys.\ Rev.\ D {\bf 68}, 014507 (2003);
	 R.V.~Gavai and S.~Gupta,
	 Phys.\ Rev.\ D {\bf 68}, 034506 (2003);
	 C.R.~Allton, M.~D\"{o}ring, S.~Ejiri, S.J.~Hands, O.~Kaczmarek,
	 F.~Karsch, E.~Laermann and K.~Redlich
	 %``Thermodynamics of two flavor QCD to sixth order in quark chemical potential,''
	 Phys.\ Rev.\  D{\bf 71}, 054508 (2005).
	 
 \bibitem{bazavov}
	 A.~Bazavov and P.~Petreczky,
	 %``Deconfinement and chiral transition with the highly improved staggered
	 %quark (HISQ) action,''
	 J.~Phys.~Conf.~Ser. \textbf{230} 012014 (2010).

 \bibitem{fodor}
	 S.~Borsanyi, Z.~Fodor, C.~Hoelbling, S.~D.~Katz, S.~Krieg,
	 C.~Ratti and K.~K.~Szabo
	 [Wuppertal-Budapest Collaboration],
	 %``Is there still any Tc mystery in lattice QCD? Results with physical masses
	 %in the continuum limit III,''
	 %arXiv:1005.3508 [hep-lat].
	 JHEP \textbf{1009}, 073 (2009);  S.~Borsanyi, Z.~Fodor,
	 S.~D.~Katz, S.~Krieg, C.~Ratti and K.~Szabo,
	 %``Fluctuations of conserved charges at finite temperature from lattice QCD,''
	 JHEP {\bf 1201}, 138 (2012)
	 %	 [arXiv:1112.4416 [hep-lat]].
	 %%CITATION = ARXIV:1112.4416;%%
	

 \bibitem{lgtl}
	 A.~Bazavov {\it et al.}  [HotQCD Collaboration],
	 %``Fluctuations and Correlations of net baryon number, electric
	 %charge, and strangeness: A comparison of lattice QCD results
	 %with the hadron resonance gas model,''
	 Phys.~Rev.~D \textbf{86}, 034509 (2012).
	 % [arXiv:1203.0784 [hep-lat]].

 \bibitem{model1}
	 B.~Stokic, B.~Friman and K.~Redlich,
	 %``Kurtosis and compressibility near the chiral phase transition,''
	 Phys.\ Lett.\ B {\bf 673}, 192 (2009);
	 V.~Skokov, B.~Stokic, B.~Friman and K.~Redlich,
	 %``Meson fluctuations and thermodynamics of the Polyakov loop extended quark-meson model,''
	 Phys.\ Rev.\ C {\bf 82}, 015206 (2010).
	 
 \bibitem{model2}
	 C.~Sasaki, B.~Friman and K.~Redlich,
	 %``Density Fluctuations as Signature of a Non-Equilibrium First Order Phase Transition,''
	 J.\ Phys.\ G {\bf 35}, 104095 (2008);
	 C.~Sasaki, B.~Friman and K.~Redlich,
	 %``Susceptibilities and the Phase Structure of a Chiral Model with Polyakov Loops,''
	 Phys.\ Rev.\ D {\bf 75}, 074013 (2007);
	 C.~Sasaki, B.~Friman and K.~Redlich,
	 %``Quark Number Fluctuations in a Chiral Model at Finite Baryon Chemical Potential,''
	 Phys.\ Rev.\ D {\bf 75}, 054026 (2007).
	 
 \bibitem{model3}
	 M.~Asakawa, S.~Ejiri and M.~Kitazawa,
	 %``Third moments of conserved charges as probes of QCD phase structure,''
	 Phys.\ Rev.\ Lett.\  {\bf 103}, 262301 (2009).
	 
 \bibitem{model4}
	 T.~K.~Herbst, J.~M.~Pawlowski and B.~-J.~Schaefer,
	 %``The Impact of Fluctuations on QCD Matter,''
	 Acta.\ Phys.\ Pol. B (Proc. Suppl.) \textbf{5}, 733 (2012),
%	 [arXiv:1202.0758 [hep-ph]].
	
 \bibitem{model5}
	 B.-J.~Schaefer and M.~Wagner, Phys.\ Rev.\ D {\bf 85}, 034027 (2012)

 \bibitem{Aoki:2012yj}
	 S.~Aoki, H.~Fukaya and Y.~Taniguchi,
	 %``Chiral symmetry restoration, eigenvalue density of Dirac operator and axial U(1) anomaly at finite temperature,''
	 Phys.\ Rev.\ D {\bf 86}, 114512 (2012)
	 %	 [arXiv:1209.2061 [hep-lat]].
	 %%CITATION = ARXIV:1209.2061;%%
	 %9 citations counted in INSPIRE as of 17 May 2013

 \bibitem{pn_qm_frg}	 
	 K.~Morita, B.~Friman, K.~Redlich and V.~Skokov, Phys.\ Rev.\ C
	 {\bf 88}, 034903 (2013).

 % \bibitem{Lucasc}
	 %	 E. Lukacs,  Characteristic functions,  London, Griffin (1970).
	 
 \bibitem{rw}
	 A.~Roberge and N.~Weiss, Nucl. Phys. {\bf B275}, 734 (1986).

 \bibitem{immu1}
	 N.~Weise, Phys.\ Rev.\ D {\bf 35}, 2495 (1987).

 \bibitem{immu2}
	 P.~de Forcrand and O.~Philipsen, Nucl.\ Phys. {\bf B642}, 290
	 (2002); M. ~D'Elia and F.~Sanfilippo, Phys.\ Rev.\ D{\bf 80},
	 111501(R), (2009).
	
 \bibitem{immu3}	
	 Y.~Sakai, K.~Kashiwa, H.Kouno, and M.~Yahiro, Phys.\ Rev.\ D{\bf
	 77}, 051901(R) (2008).

 \bibitem{morita}
	 K.~Morita, V.~Skokov, B.~Friman and K.~Redlich,
	 %``Probing deconfinement in a chiral effective model with Polyakov loop at imaginary chemical potential,''
	 Phys.\ Rev.\ D {\bf 84}, 076009 (2011); Phys.\ Rev.\ D {\bf 84},
	 074020 (2011); Acta.\ Phys.\ Pol. B (Proc. Suppl.) \textbf{5}, 803
	 (2012).
	 %[arXiv:1111.3446 [hep-ph]].

 \bibitem{Engels-Karsch}
	 J.~Engels and F.~Karsch, Phys.\ Rev.\ D \textbf{85}, 094506
	 (2012).

 \bibitem{hwa}
	 P. Braun-Munzinger, K. Redlich, J. Stachel, in Quark-Gluon Plasma
	 3, Eds. R.C. Hwa and X.N. Wang, (World Scientific Publishing, 2004).
	 
 \bibitem{cleymans1}
	 J.~Cleymans, P.~Koch, Z. Phys. C {\bf 52}, 137 (1991);\\
	 C.~M.~Ko, V.~Koch, Z.-W.~Lin, K.~Redlich, M.~A.~Stephanov, and
	 X.-N.~Wang,  Phys.\ Rev.\ Lett. {\bf 86}, 5438 (2001).
	 
 \bibitem{turko}
	 J.~Cleymans, K.~Redlich and L.~Turko,
	 %``Probability distributions in statistical ensembles with conserved charges,''
	 Phys.\ Rev.\  C {\bf 71}, 047902 (2005).
	 
 \bibitem{hagedorn}
	R.~Hagedorn and K.~Redlich,
	%``Statistical Thermodynamics In Relativistic Particle And Ion Physics: Canonical Or Grand Canonical?,''
	Z.\ Phys.\ C{\bf 27}, 541 (1985).
	
 \bibitem{stephanov06:_qcd_critic_point_and_compl}
	 M.~A.~Stephanov,
	 Phys.\ Rev.\ D \textbf{73}, 094508 (2006).
	 
 \bibitem{alford99:_imagin_chemic_poten_and_finit}
	 M.~Alford, A.~Kapustin and F.~Wilczek,
	 Phys.\ Rev.\ D \textbf{59}, 054502 (1999).
	 
 \bibitem{forcrand06:_finit_densit_qcd_with_canon_approac}
	 P.~de~Forcrand and S.~Kratochvila,
	 Nucl.\ Phys.\ B. (Proc. Suppl.) \textbf{153}, 62 (2006).
	 
 \bibitem{ejiri08:_canon_qcd}
	 S.~Ejiri, 
	 Phys.\ Rev.\ D \textbf{78}, 074507 (2008).
		
 \bibitem{li:_finit_qcd_n_n}
	 A.~Li, A.~Alexandru, K.~F.~Liu and X.~Meng,
	 Phys.\ Rev.\ D \textbf{82}, 054502 (2010).
			
 \bibitem{li11:_critic_point_of_n_f}
	 A.~Li, A.~Alexandru and K.-F.~Liu,
	 Phys.\ Rev.\ D \textbf{84}, 071503(R)(2011).
	
 \bibitem{XQCDJ}
	 K.~Nagata, S.~Motoki, Y.~Nakagawa, A.~Nakamura, and T.~Saito (XQCD-J
	 Collaboration), Prog.\ Theor.\ Exp.\ Phys.\ 01A103 (2012).
	
 \bibitem{Yang:1952be}
	 C.~-N.~Yang and T.~D.~Lee,
	 %``Statistical theory of equations of state and phase transitions. 1. Theory of condensation,''
	 Phys.\ Rev.\  {\bf 87}, 404 (1952).
	 %%CITATION = PHRVA,87,404;%%
	
 \bibitem{skokov11:_mappin}
	 V.~Skokov, K.~Morita and B.~Friman,
	 Phys.\ Rev.\ D \textbf{83}, 071502(R) (2011).

 \bibitem{Friman:2012gg}
	 B.~Friman,  Acta.\ Phys.\ Pol.\ B (Proc Suppl.) \textbf{5}, 707 (2012),
%	 arXiv:1202.0021 [hep-ph].



 \bibitem{cleymans2}
	 P.~Braun-Munzinger, J.~Cleymans, H.~Oeschler and K.~Redlich,
	 %``Maximum relative strangeness content in heavy ion collisions around 30-GeV/A,''
	 Nucl.\ Phys.\ {\bf A697}, 902 (2002).
\end{thebibliography}


\end{document}